\documentclass[a4paper]{article}

\usepackage{INTERSPEECH2021}

\usepackage{graphicx}
\usepackage[dvipsnames,table]{xcolor}

\usepackage{array}
\newcolumntype{$}{>{\global\let\currentrowstyle\relax}}
\newcolumntype{^}{>{\currentrowstyle}}
\newcommand{\rowstyle}[1]{\gdef\currentrowstyle{#1}%
	#1\ignorespaces
}

\usepackage{booktabs}
\usepackage{multirow}
\usepackage{url}

\title{Speaker-conversation factorial designs for diarization error analysis}
\name{Scott Seyfarth, Sundararajan Srinivasan, Katrin Kirchhoff\thanks{The first two authors contributed equally.}}
\address{Amazon AWS AI}
\email{seyfar@amazon.com, sundarsr@amazon.com, katrinki@amazon.com}

\begin{document}

\maketitle

\begin{abstract}
Speaker diarization accuracy can be affected by both acoustics and conversation characteristics.
Determining the cause of diarization errors is difficult because speaker voice acoustics and conversation structure co-vary, and the interactions between acoustics, conversational structure, and diarization accuracy are complex.
This paper proposes a methodology that can distinguish independent marginal effects of acoustic and conversation characteristics on diarization accuracy by remixing conversations in a factorial design.
As an illustration, this approach is used to investigate gender-related and language-related accuracy differences with three diarization systems: a baseline system using subsegment x-vector clustering, a variant of it with shorter subsegments, and a third system based on a Bayesian hidden Markov model.
Our analysis shows large accuracy disparities for the baseline system primarily due to conversational structure, which are partially mitigated in the other two systems.
The illustration thus demonstrates how the methodology can be used to identify and guide diarization model improvements.
\end{abstract}

\section{Introduction}
\label{sec:introduction}

Diarization is the process of automatically clustering spoken utterances into groups such that each group contains all and only the utterances produced by a single speaker.
Automatic diarization systems are considered to be more accurate when they create groups of utterances that are closer to this ideal.
What affects the accuracy of automatic diarization?
Accuracy is logically influenced by \emph{acoustics}: for instance, accuracy will be worse if a system cannot easily differentiate the voices of two speakers.
However, \emph{conversation structure}---the number of speakers and the schedule of their turns and pauses---also affects accuracy.
Most diarization systems operate by extracting and clustering short speech subsegments \cite{garcia-romero_speaker_2017} (see \cite{park_review_2021} for a review of alternatives), which is most reliable when those subsegments do not span across speaker turn boundaries and when the clusters are similar in size.
As a consequence, these systems have better accuracy in balanced conversations with long turns and little turn overlap \cite{knox_where_2012,mirghafori_nuts_2006}.

Distinguishing the two sources of error is difficult because groups of speakers with different voice acoustics may also engage in conversation differently.
For example, previous work has documented gendered differences in turn-taking style \cite{edelsky1981,tannen_turn-taking_2012}, which implies that differences in diarization accuracy for female and male speakers could be due to turn-taking structure as well as acoustics.
While a regression could be used to distinguish these variables, such an analysis is only realistic with a prior specification of how the relevant conversational and acoustic variables interact to affect accuracy.

We propose analyzing diarization error using an experimental intervention that de-correlates speaker voice characteristics and conversation structures in the experimental design rather than in the analysis.
We describe a procedure for remixing conversations in a speaker-conversation factorial design.
We illustrate the procedure by generating experiments based on two widely-used corpora, and use the experiments to test three open-source diarization systems: a baseline system that uses x-vector-based AHC; a variant of the same system with shorter subsegments; and a third system with Bayesian HMM frame-realignment.
We analyze the errors that they produce, which indicate specific directions for improvement.

\section{Method}
\label{sec:method}
\label{sec:method:factorial}

Consider the following structure with seven speech segments and two roles, A and B. One speaker (in dark purple) produces three consecutive segments, and then a second speaker (in light green) asks a question over two segments, then the purple speaker responds with two segments.

{\centering
	\vspace{4pt}
	\begin{tabular}{$l^c^c^c^c^c^c^c}
	\rowstyle{\bfseries}
	Structure: & \textcolor{Plum}{A} & \textcolor{Plum}{A} & \textcolor{Plum}{A} & \textcolor{YellowGreen}{B} & \textcolor{YellowGreen}{B} & \textcolor{Plum}{A} & \textcolor{Plum}{A} \\
	{\itshape\small Duration (sec)}: & 3.0 & 2.0 & 4.0 & 1.5 & 0.5 & 6.0 & 2.0 \\
	\end{tabular}
}

If a diarization system attributes the first B segment to A, the F1 scores for the speakers in roles A and B are 0.96 and 0.40, respectively. Is the difference because the diarization system has difficulty with the green speaker's voice characteristics, or is it because role B has less speech than role A to use for cluster estimation? To answer this question, one can test mirror-image versions of the structure in which both speakers are assigned to both possible roles:

{\centering\bfseries
	\vspace{2pt}
	\begin{tabular}{l c c c c c c c}
		Version 1: & \textcolor{Plum}{A} & \textcolor{Plum}{A} & \textcolor{Plum}{A} & \textcolor{YellowGreen}{B} & \textcolor{YellowGreen}{B} & \textcolor{Plum}{A} & \textcolor{Plum}{A} \\
		Version 2: & \textcolor{YellowGreen}{A} & \textcolor{YellowGreen}{A} & \textcolor{YellowGreen}{A} & \textcolor{Plum}{B} & \textcolor{Plum}{B} & \textcolor{YellowGreen}{A} & \textcolor{YellowGreen}{A} \\
	\end{tabular}
	\vspace{2pt}
}

By testing both possible versions, the effects of speaker voice characteristics can be separated from the effects of the conversation structure. The two versions will not occur in a natural speech corpus. Instead of using natural conversations, both versions are constructed by splicing source audio from the two speakers according to the desired structure \cite{karpov_free_2018}. If there are more than two speakers and roles, then all factorial combinations of speakers and roles can be generated.

After simulating and testing both versions of the conversation, an F1 score (or any other class-specific metric) is calculated for each speaker and for each role, averaging over the two versions.
If the green speaker has significantly lower average F1, the errors must be partially due to characteristics involving that speaker's voice or channel.
Similarly, if role B has lower F1 than role A regardless of the actual speakers assigned to each role, then at least part of the diarization error is due to that role.
Alternatively, diarization error rate (DER) can be calculated for both versions of this structure and compared to the factorial simulations of another structure using the same two speaker voices.
If DER is still worse for this structure even with the same two voices, then the structure itself must be a source of higher error.

\section{Experiment 1: Gender in Fisher}
\label{sec:gender}

To illustrate the proposed technique, we first evaluate how gender correlates with the accuracy of three version of an example model for American English. Gender is defined here as the apparent female or male gender perceived by an annotator; our analysis thus evaluates only the perception of gender without allocating gender experience or identity to voices.
Female and male voices can have different acoustic properties due to physical differences in vocal tract anatomy \cite{kahane1978,hanson_glottal_1999} and socialized norms \cite{van_bezooijen_sociocultural_1995}. Therefore, gender might correlate with diarization accuracy if the embedding model has more difficulty with some acoustic features (e.g., wide formant spacing) than others.
At the same time, gender can also be associated with differences in turn-taking style \cite{edelsky1981,tannen_turn-taking_2012,anderson_meta-analyses_1998}, and these differences might also produce a correlation between gender and diarization accuracy.
If a system developer wants to achieve equal performance across groups, these two possibilities must be distinguished to identify the system component that needs to be improved.

\subsection{Example diarization systems}
\label{sec:system}

For the illustration, we evaluated three versions of the 8 kHz diarization system from the Kaldi CALLHOME diarization v2 recipe \cite{povey2011kaldi}, a standard diarization benchmark. In the baseline system, speaker embeddings were extracted using 1.5 second audio subsegments with 0.75 second overlap, then grouped together with agglomerative hierarchical clustering using PLDA scores \cite{prince_probabilistic_2007} for the similarity matrix. The second system had shorter 0.75-second subsegments; because shorter subsegments imply that a smaller proportion will intersect speaker turn boundaries, we hypothesized that this will reduce sensitivity to any differences in turn-taking structure. The third system (`VB') included frame-level resegmentation with a Bayesian HMM as described in \cite{diez_speaker_2018}.

\subsection{Evaluation data}
\label{sec:evaluation-data}

The systems were evaluated on the Fisher Corpus of American English \cite{cieri2004fisher} (LDC2004S13, LDC2005S13, LDC2004T19, LDC2005T19), which contains 10-minute partial transcriptions of 11,699 two-person conversations, including 4,736 with two female speakers, 3,143 with two male speakers, and 3,820 with one female and one male speaker.
Conversation style is naturally individual and context-dependent, and our analysis and conclusions are limited by design only to aggregated differences in this specific corpus of conversations.

We initially tested the diarization systems on the original Fisher Corpus. The baseline system had significantly higher median DER (omitting overlapped speech) for the conversations with two female speakers (3.49) than the conversations with two male speakers (3.02) or mixed-gender conversations (2.83; all pairwise differences $p < 0.0001$ by Mann-Whitney test). This pattern also held for the VB system (2.79, 2.40, and 2.57 DER respectively), but was \emph{reversed} for the system with 0.75s subsegments (3.83, 4.17, 2.73). As suggested above, the DER differences might be due to gendered voice acoustics, conversation structure (e.g., the rate of turn changes), or both.

\subsection{Conversation structures and speaker pairs}
\label{sec:schemas}

To separate the effects of conversation structures from speaker voices, the systems were evaluated on a version of the corpus constructed using the procedure in \S\ref{sec:method:factorial}. One hundred fifty conversations were drawn randomly from Fisher, including 50 in which the original two speakers were both female, 50 in which the original two speakers were both male, and 50 mixed-gender conversations. For each conversation, a structure was extracted by taking the original sequence of segments, with each segment labeled with its duration and a speaker role (A, B, or silence), with no turn overlap.

Independently, we drew a random sample of 300 speakers with at least 2 minutes of audio each. We selected American English speakers raised in California who were ages 20--30. We chose this demographic to maximize dialect homogeneity given the limited (meta)data in Fisher. This naturally limits our conclusions to this demographic within this corpus. These speakers were randomly re-paired into 50 same-gender female pairs, 50 same-gender male pairs, and 50 mixed-gender pairs.

\subsection{Conversation design}
\label{sec:schema-versions}

A simulated version of Fisher was created based on the samples of conversation structures and speaker pairs. The three kinds of structures (same-gender female, same-gender male, and mixed-gender) were crossed with the three kinds of speaker pairs (same-gender female, same-gender male, and mixed) to guarantee independence of conversation structure and speaker gender. For every unique combination of structure and speaker pair, two mirror versions of the conversation were created by assigning each speaker in the pair to each of the two roles in the structure, as described in \S\ref{sec:method:factorial}.

The original audio of the transcribed utterances for each speaker was extracted. Each transcribed utterance was re-seg\-men\-ted to remove silences and breath noises using a pre-trained voice activity detection model from \cite{Bredin2020}, then concatenated together into one continuous speech stream per speaker. The segments in each conversation version were then filled in by copying audio from the appropriate speaker's audio stream for the specified segment duration, with a 0.01 second volume taper at the beginning and end of each segment to avoid audio artifacts. If one or both speakers did not have enough original audio to fill all of their assigned segments in a structure, the structure was truncated to the maximum length such that both speakers could take both roles.
This produced 45,000 simulated conversations (150 pairs $\times$ 150 structures $\times$ 2 versions), each 2--13 minutes

\subsection{Scoring}
\label{sec:scoring}

The systems were tested on the simulated conversations and scored using the US NIST SCTK \texttt{md-eval.pl} script with no forgiveness collar around speaker boundaries.
To focus only on between-speaker confusions involving the core diarization system, we provided oracle segmentation \cite{huijbregts_speaker_2012,garcia-romero_speaker_2017,diez_speaker_2018}.

Per-speaker F1 scores were calculated for each conversation assuming optimal mappings between the reference and hypothesized system speakers.
The scores for each speaker were then averaged over the two versions of each conversation that they they participated in (see \S\ref{sec:method:factorial}), resulting in one F1 score per speaker per pair per structure.
We analyze F1 scores here, rather than DER, so that we can test the difference in accuracy between the female and male speaker within a mixed-gender conversation (see \S\ref{sec:method}).

\subsection{Results}

\begin{figure}
	 \centering
	 \includegraphics[width=\columnwidth]{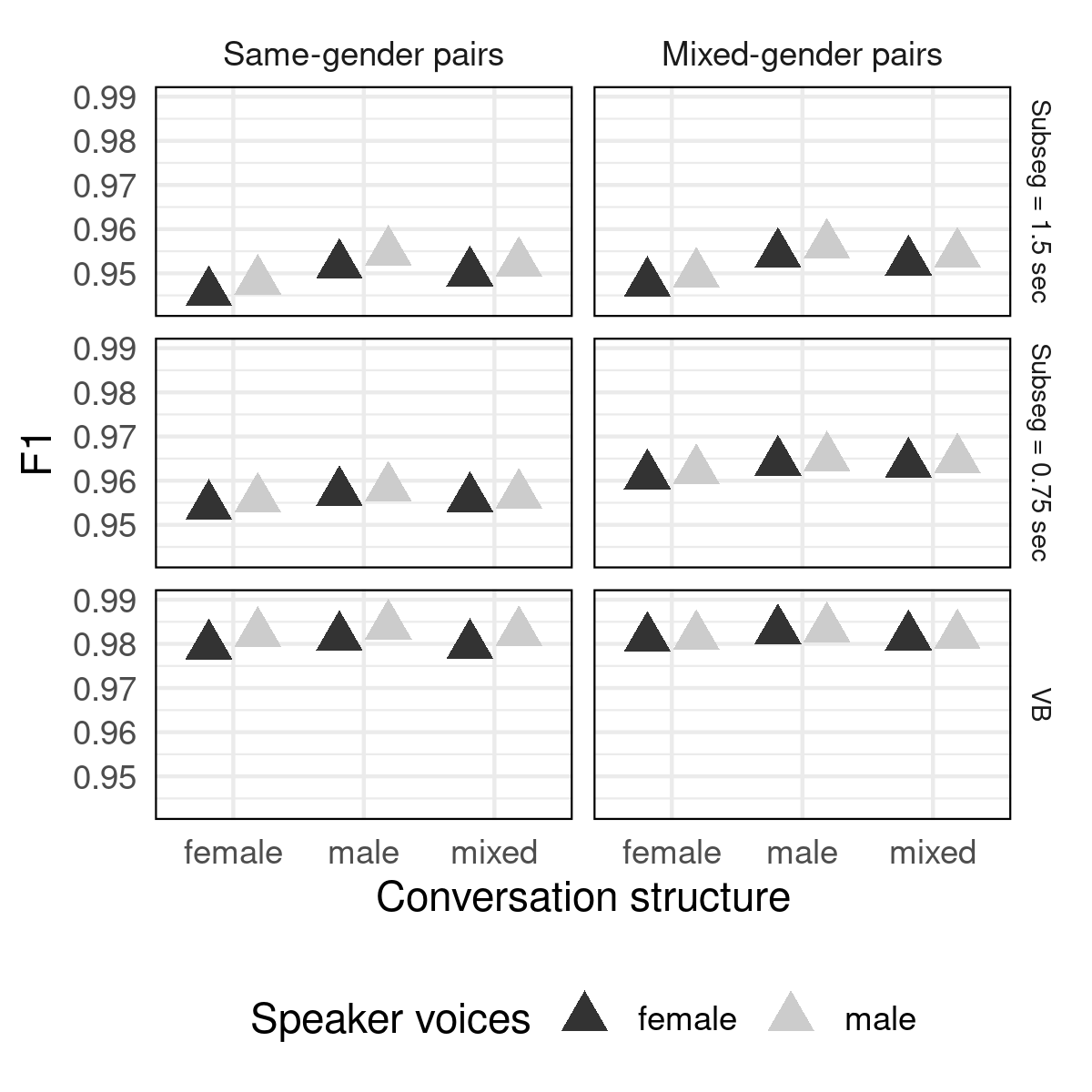}
	 \caption{Median F1 scores for each combination of speaker voice gender, structure gender, and speaker pair gender.}
	 \label{figure:fisher-f1}
\end{figure}

Figure \ref{figure:fisher-f1} shows median speaker F1 in the simulated corpus.
To test the accuracy differences to perceived speaker gender, the median F1 was calculated for each of the speaker pairs, averaging over the 150 structures. Median F1 was lower for the 50 female pairs than the 50 male pairs for all three systems, even though all of the speaker pairs had been remixed into the same conversation. This difference is significant for the baseline system ($\Delta\mbox{F1}=0.003$, $p = 0.029$ by Mann-Whitney test) and the VB system ($\Delta0.002$, $p = 0.0002$), but not the 0.75s subsegment system ($\Delta0.0007$, $p = 0.410$).

Additionally, the median F1 was calculated for the female voices in the mixed-gender pairs and compared to the male voices in the mixed-gender pairs. This difference was not significant ($p = 0.11$, $0.15$, $0.34$). Thus, while there was evidence that accuracy is different for female and male voices in same-gender pairs for 2/3 systems, there was no evidence for a difference in mixed-gender pairs.

Next, the median F1 was calculated for each of the 150 structures, averaging over all 150 speaker pairs. Median F1 was lower when the \emph{original} conversation had two female speakers than when it had had two male speakers, regardless of the speaker voices that had been remixed into the conversation. The difference was significant for the baseline system ($\Delta0.011$; $p = 0.016$) and the VB system ($\Delta0.001$; $p = 0.049$), but not the 0.75s subsegment system ($\Delta0.003$; $p = 0.155$). Notably, the baseline system with long subsegments has an F1 loss associated with conversation structure that is an order of magnitude larger than (i) the loss associated with speaker voices and (ii) the loss for the short-subsegment system, as predicted in \S\ref{sec:system}.

\section{Experiment 2: CALLHOME languages}
\label{sec:experiment-2}

Diarization systems are often used for multiple languages. As with gender, both acoustics and differences in turn-taking style might lead to accuracy differences between languages. The procedure was the same as \S\ref{sec:gender} except as described below.

\subsection{Evaluation data}
\label{sec:callhome-system}
\label{sec:callhome-evaluation}

The three systems described in \S\ref{sec:system} were evaluated on the language-specific CALLHOME-6 datasets (LDC97S42--43, LDC97S45, LDC96S34--35, LDC96S37, LDC97T14--19), which include partial transcriptions of 100--120 conversations for six languages.

The upper panel of Table \ref{table:callhome-der} shows DER for the subset of two-speaker conversations.
DER is substantially worse for the Japanese set, which could be due to differences in conversation style. For example, \cite{maynard_conversation_1990,Stivers10587} suggest that Japanese conversations involve shorter gaps between utterances and substantially more short backchannels than American English ones. Both patterns could reduce accuracy due to more speech subsegments that include multiple speakers (\S\ref{sec:system}), which will interfere with cluster estimation even when these subsegments are not scored.

To quantify this, the lower panel of the table shows the average speaker entropy during subsegments of 1.5 seconds and 0.75 seconds within each set. Values closer to 0 means that most subsegments contained only one speaker; closer to 1 means that most subsegments were evenly divided between two speakers. The Japanese set has substantially higher entropy than the other sets, though shorter subsegments reduce the differences.

However, higher DER could also be caused by speaker embedding biases \cite{liang_lu_effect_2009,xia_cross-lingual_2019}, as only two of 12 corpora in the embedding model training set \cite{snyder2018xvector} have any Japanese speech (LDC2011S05, LDC2011S08), and these two include less than 2\% Japanese conversations.
To evaluate whether the differences in DER among the six datasets can be attributed to acoustic features or to differences in conversational style, we simulated conversations in which the speaker audio comes from one language set (e.g., Japanese) and the conversation structure comes from another language set (e.g., German), for all factorial combinations of audio and structure language.

\begin{table}
	\caption{Median DER aggregated by conversation with overlapped speech omitted (upper panel) and subsegment speaker entropy (lower panel) for each original CALLHOME-6 subset. Key: \textbf{ar}: Egyptian Arabic (70 conversations), \textbf{en} = American English (109), \textbf{de} = German (82), \textbf{jp} = Japanese (99), \textbf{zh} = Mandarin (57), \textbf{es} = Spanish (71).\label{table:callhome-der}}
	\centering
	
	\begin{tabular}{l r r r r r r}
		\toprule
		& ar & en & de & jp & zh & es \\
		\midrule
		1.50 sec & $7.48$ & $6.51$ & $5.79$ & $12.90$ & $6.69$  & $6.39$ \\
		0.75 sec & $7.72$ & $6.44$ & $7.27$ & $12.60$ & $6.60$  & $7.73$ \\
		VB & $5.41$ & $4.93$ & $4.60$ & $9.05$ & $4.86$  & $5.14$ \\
		\midrule
		\midrule
		1.50 sec & $0.10$ & $0.07$ & $0.10$ & $0.17$ & $0.07$  & $0.11$ \\
		0.75 sec & $0.07$ & $0.06$ & $0.07$ & $0.12$ & $0.06$  & $0.07$ \\
		\bottomrule
	\end{tabular}
	
\end{table}

\subsection{Sample and experiment design}

For each language set, we selected a random sample of 20 pairs of female speakers, 20 pairs of male speakers, and 20 mixed-gender pairs from the original corpus.
We selected the original speaker pairs rather than re-pairing speakers as in Experiment 1, which ensures that speakers within each language are uniformly selected from the at-home and abroad sets, and only selected pairs where both were listed as having the same accent to help minimize dialect differences.
For some sets, there were $<$20 pairs per gender group meeting the criteria.

The conversation structure and speaker audio was extracted for each pair, following the procedure used for Fisher in Experiment 1. Each conversation structure was rebuilt with spliced audio from every speaker pair of the same gender across the six languages. For example, the 20 conversation structures from the female Egyptian Arabic pairs were each simulated using the audio from every female pair in all six languages.
This produced $\sim$86,400 simulated conversation versions ([$\sim$20 pairs $\times$ 6 languages] $\times$ [$\sim$20 structures $\times$ 6 languages] $\times$ 3 genders $\times$ 2 versions), each 1--13 minutes.

\subsection{Results}
\label{sec:callhome-scoring}

\begin{figure}
	\centering
	\includegraphics[width=\columnwidth]{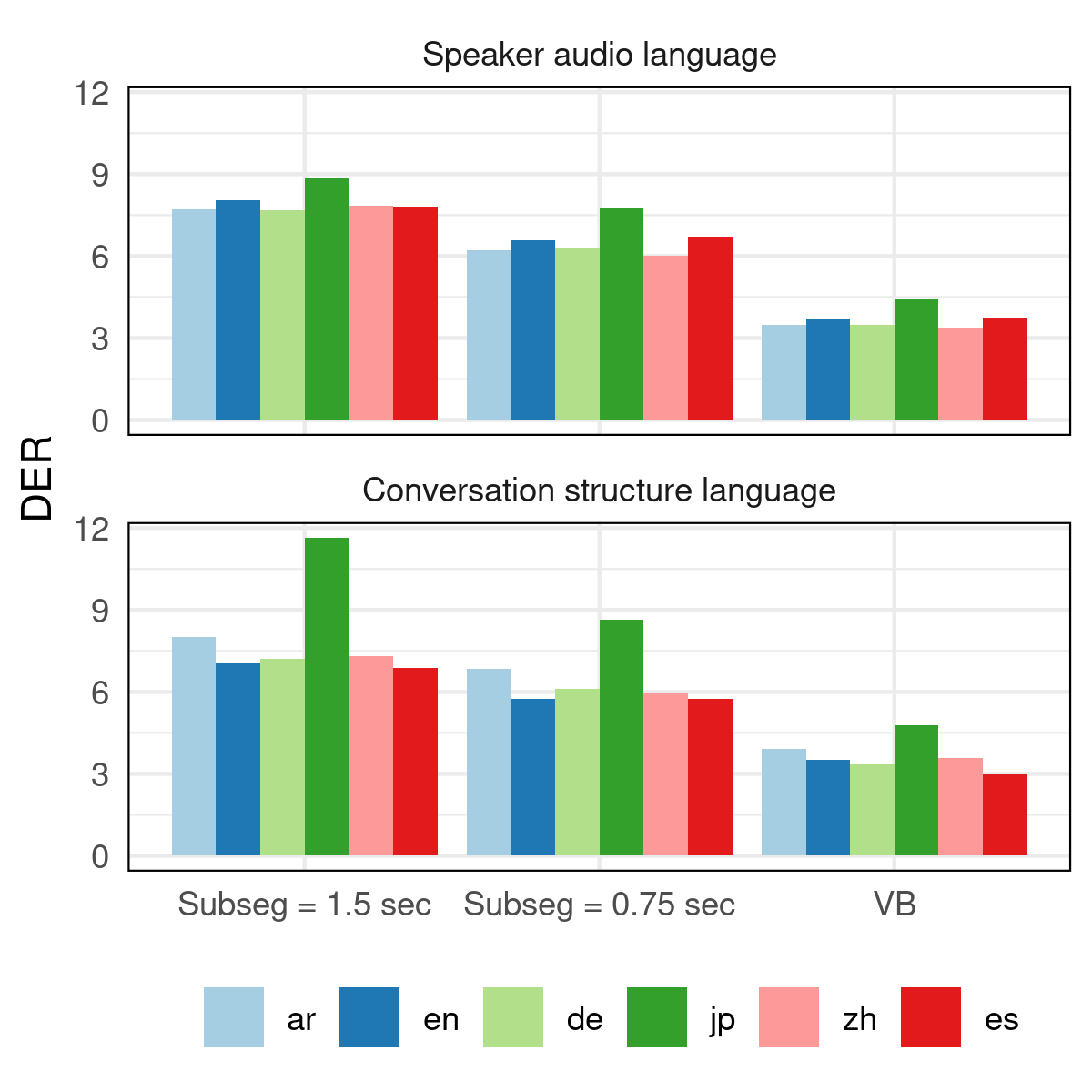}
	\caption{Median DERs for speaker pairs (upper) and conversation structures (lower) in the simulated CALLHOME-6 corpus, marginalizing over structures (upper) or speaker pairs (lower), with the six languages weighted equally. Key in Table \ref{table:callhome-der}.\label{figure:callhome-der}}
\end{figure}

Figure \ref{figure:callhome-der} shows median DERs for the speaker pairs from each language (upper panel; averaging over all conversation structures)
and for the conversation structures from each language (lower panel; averaging over all speaker pairs).
Notably, the Japanese structures have substantially higher DER than structures from the other languages, particularly for the baseline model.
However, when the Japanese speaker pairs are remixed into the same conversation structures as the other speakers, DER is not much higher for the Japanese speakers than the others.

To test these differences, we first calculated the median DER for each speaker pair separately for each model, averaging over all of the structures. We fit a quantile regression to predict median DER from the speakers' language, gender, and the model. We calculated the marginal posterior differences in DER between the speakers' languages with 95\% Bayesian credible intervals (CIs) using \cite{burkner2018,JSSv076i01,lenth2020}.
When conversation structure is held constant, the estimated differences between the speakers' languages are small, at most $\Delta$DER $=0.53$ between languages for the baseline model, $\Delta0.69$ for the 0.75s subsegment model, and $\Delta0.44$ for the VB model.

Next, we calculated median DER for each conversation structure, averaging over the speaker pairs from all of the languages. We fit a quantile regression to predict DER from the original language of the conversation structure, the original gender, and the model, following the same procedure. The estimated marginal differences between the structures are much larger, up to $\Delta$DER $=4.99$ for the baseline model, $\Delta2.99$ for the 0.75s subsegment model, and $\Delta1.60$ DER for the VB model. The largest pairwise differences all involved Japanese (baseline: 4.20--4.99 DER; 0.75s: 2.19--2.99; VB: 0.90--1.60), and all CIs involving Japanese structures indicated 95\%+ probability that the Japanese structures had greater DER than others.

The small marginal DER differences with respect to the audio, combined with the large differences with respect to conversation structure, together indicate that the DER differences between the CALLHOME-6 language sets for this model are driven mostly by different turn-taking styles, not acoustics.

\section{Discussion}
\label{sec:discussion}

We proposed simulating conversations in a factorial design that makes conversational structure independent of speaker voice characteristics, and showed that a baseline diarization model has better accuracy for both the male voices and male conversation structures than the female ones in the Fisher Corpus.
The differences between female and male conversation structures were much larger than the marginal differences between female and male voices.
We also found that differences in conversation structure in the CALLHOME-6 sets are associated with large differences in diarization accuracy (up to 4.99 DER), but that differences in accuracy between the language sets are small (up to 0.69 DER) when the conversation structure is held constant.

For the Fisher conversations, reducing the subsegment length from 1.5s to 0.75s shrank the marginal accuracy differences between the female and male conversation structures by an order of magnitude, as did training directly on the distribution of speaker changes (VB; though this did not improve accuracy differences due to voices).
Similarly, the Japanese conversation structures were associated with significantly higher DER than the other conversation structures, but this was improved with shorter subsegments and HMM sequence training.

By design, our study is constrained to estimating marginal outcomes for a subset of the gender and language groups in two specific datasets.
Individuals vary substantially in conversational style, and gendered and language-related styles will vary across other conversational contexts.
More generally, though, models which assume that input speech subsegments do not span turn boundaries will have lower performance for speakers that take shorter turns, backchannel more, or have shorter gaps between turns \cite{knox_where_2012,mirghafori_nuts_2006}.
As we illustrated, these disparities can be partially mitigated by using shorter subsegments, which leads to a lower proportion of subsegments that span turn boundaries (Table \ref{table:callhome-der}, lower panel).
This highlights the importance of turn-taking structure in understanding and improving diarization error profiles: if a model is not trained or tuned on conversation structure, structure-related accuracy problems cannot be improved with more or better training data.
Model designs such as UIS-RNN \cite{zhang_fully_2019} and end-to-end sequence models \cite{park_review_2021,fujita_end--end_2019,fujita_end--end_20192} thus offer more opportunities to reduce between-group accuracy disparities with careful data selection.

Our illustrated example focused on cluster assignment errors produced by a subsegment-based model.
However, the method that we proposed can be applied to analyze the error profiles of any class of diarization model, including errors due to segmentation or other system components \cite{knox_where_2012,huijbregts_speaker_2012}.
Because the method requires only a specially-constructed dataset, it can be used equally to evaluate other diarization components \cite{huijbregts_blame_2007} or end-to-end systems that are otherwise resistant to introspection.
Future work might also further explore the relationship of conversation characteristics with accuracy by manipulating specific conversation characteristics in synthetic structures, such as the rate of turn changes, amount of speaker overlap, and number of speakers \cite{knox_where_2012,mirghafori_nuts_2006,karpov_free_2018}.

\bibliographystyle{IEEEtran}
\bibliography{refs}

\begin{thebibliography}{10}
\providecommand{\url}[1]{#1}
\csname url@samestyle\endcsname
\providecommand{\newblock}{\relax}
\providecommand{\bibinfo}[2]{#2}
\providecommand{\BIBentrySTDinterwordspacing}{\spaceskip=0pt\relax}
\providecommand{\BIBentryALTinterwordstretchfactor}{4}
\providecommand{\BIBentryALTinterwordspacing}{\spaceskip=\fontdimen2\font plus
\BIBentryALTinterwordstretchfactor\fontdimen3\font minus
  \fontdimen4\font\relax}
\providecommand{\BIBforeignlanguage}[2]{{%
\expandafter\ifx\csname l@#1\endcsname\relax
\typeout{** WARNING: IEEEtran.bst: No hyphenation pattern has been}%
\typeout{** loaded for the language `#1'. Using the pattern for}%
\typeout{** the default language instead.}%
\else
\language=\csname l@#1\endcsname
\fi
#2}}
\providecommand{\BIBdecl}{\relax}
\BIBdecl

\bibitem{garcia-romero_speaker_2017}
\BIBentryALTinterwordspacing
D.~Garcia-Romero, D.~Snyder, G.~Sell, D.~Povey, and A.~McCree,
  ``\BIBforeignlanguage{en}{Speaker diarization using deep neural network
  embeddings},'' in \emph{\BIBforeignlanguage{en}{2017 {IEEE} {International}
  {Conference} on {Acoustics}, {Speech} and {Signal} {Processing}
  ({ICASSP})}}.\hskip 1em plus 0.5em minus 0.4em\relax New Orleans, LA: IEEE,
  Mar. 2017, pp. 4930--4934. [Online]. Available:
  \url{http://ieeexplore.ieee.org/document/7953094/}
\BIBentrySTDinterwordspacing

\bibitem{park_review_2021}
\BIBentryALTinterwordspacing
T.~J. Park, N.~Kanda, D.~Dimitriadis, K.~J. Han, S.~Watanabe, and S.~Narayanan,
  ``\BIBforeignlanguage{en}{A {Review} of {Speaker} {Diarization}: {Recent}
  {Advances} with {Deep} {Learning}},''
  \emph{\BIBforeignlanguage{en}{arXiv:2101.09624 [cs, eess]}}, Jan. 2021,
  arXiv: 2101.09624. [Online]. Available: \url{http://arxiv.org/abs/2101.09624}
\BIBentrySTDinterwordspacing

\bibitem{knox_where_2012}
M.~T. Knox, N.~Mirghafori, and G.~Friedland, ``\BIBforeignlanguage{en}{Where
  did {I} go wrong?: {Identifying} troublesome segments for speaker diarization
  systems},'' \emph{\BIBforeignlanguage{en}{Interspeech}}, 2012.

\bibitem{mirghafori_nuts_2006}
\BIBentryALTinterwordspacing
N.~Mirghafori and C.~Wooters, ``\BIBforeignlanguage{en}{Nuts and {Flakes}: a
  {Study} of {Data} {Characteristics} in {Speaker} {Diarization}},'' in
  \emph{\BIBforeignlanguage{en}{{ICASSP}}}, 2006. [Online]. Available:
  \url{http://ieeexplore.ieee.org/document/1660196/}
\BIBentrySTDinterwordspacing

\bibitem{edelsky1981}
C.~Edelsky, ``Who's got the floor?'' \emph{Language in Society}, vol.~10,
  no.~3, pp. 383--421, 1981.

\bibitem{tannen_turn-taking_2012}
D.~Tannen, ``Turn-taking and intercultural discourse and communication,'' in
  \emph{The {Handbook} of {International} {Discourse} and {Communication}},
  C.~B. Paulston, S.~F. Kiesling, and E.~S. Rangel, Eds.\hskip 1em plus 0.5em
  minus 0.4em\relax Blackwell Publishing Ltd., 2012, pp. 135--157.

\bibitem{karpov_free_2018}
\BIBentryALTinterwordspacing
E.~Edwards, M.~Brenndoerfer, A.~Robinson, N.~Sadoughi, G.~P. Finley,
  M.~Korenevsky, N.~Axtmann, M.~Miller, and D.~Suendermann-Oeft,
  ``\BIBforeignlanguage{en}{A {Free} {Synthetic} {Corpus} for {Speaker}
  {Diarization} {Research}},'' in \emph{\BIBforeignlanguage{en}{Speech and
  {Computer}}}, A.~Karpov, O.~Jokisch, and R.~Potapova, Eds.\hskip 1em plus
  0.5em minus 0.4em\relax Cham: Springer International Publishing, 2018, vol.
  11096, pp. 113--122, series Title: Lecture Notes in Computer Science.
  [Online]. Available:
  \url{http://link.springer.com/10.1007/978-3-319-99579-3_13}
\BIBentrySTDinterwordspacing

\bibitem{kahane1978}
\BIBentryALTinterwordspacing
J.~C. Kahane, ``A morphological study of the human prepubertal and pubertal
  larynx,'' \emph{American Journal of Anatomy}, vol. 151, no.~1, pp. 11--19,
  1978. [Online]. Available:
  \url{https://onlinelibrary.wiley.com/doi/abs/10.1002/aja.1001510103}
\BIBentrySTDinterwordspacing

\bibitem{hanson_glottal_1999}
H.~M. Hanson and E.~S. Chuang, ``\BIBforeignlanguage{en}{Glottal
  characteristics of male speakers: {Acoustic} correlates and comparison with
  female data},'' \emph{\BIBforeignlanguage{en}{Journal of the Acoustical
  Society of America}}, vol. 106, no.~2, 1999.

\bibitem{van_bezooijen_sociocultural_1995}
R.~van Bezooijen, ``Sociocultural aspects of pitch differences between
  {Japanese} and {Dutch} women,'' \emph{Language and Speech}, vol.~38, no.~3,
  pp. 253--265, 1995.

\bibitem{anderson_meta-analyses_1998}
K.~J. Anderson and C.~Leaper, ``\BIBforeignlanguage{en}{Meta-{Analyses} of
  {Gender} {Effects} on {Conversational} {Interruption}: {Who}, {What}, {When},
  {Where}, and {How}},'' \emph{\BIBforeignlanguage{en}{Sex Roles}}, vol.~39,
  no. 3-4, pp. 225--252, 1998.

\bibitem{povey2011kaldi}
D.~Povey, A.~Ghoshal, G.~Boulianne, L.~Burget, O.~Glembek, N.~Goel,
  M.~Hannemann, P.~Motlicek, Y.~Qian, P.~Schwarz \emph{et~al.}, ``The {K}aldi
  speech recognition toolkit,'' 2011.

\bibitem{prince_probabilistic_2007}
\BIBentryALTinterwordspacing
S.~J. Prince and J.~H. Elder, ``\BIBforeignlanguage{en}{Probabilistic {Linear}
  {Discriminant} {Analysis} for {Inferences} {About} {Identity}},'' in
  \emph{\BIBforeignlanguage{en}{{ICCV} 11}}, 2007. [Online]. Available:
  \url{http://ieeexplore.ieee.org/document/4409052/}
\BIBentrySTDinterwordspacing

\bibitem{diez_speaker_2018}
\BIBentryALTinterwordspacing
M.~Diez, L.~Burget, and P.~Matejka, ``\BIBforeignlanguage{en}{Speaker
  {Diarization} based on {Bayesian} {HMM} with {Eigenvoice} {Priors}},'' in
  \emph{\BIBforeignlanguage{en}{Odyssey 2018 {The} {Speaker} and {Language}
  {Recognition} {Workshop}}}.\hskip 1em plus 0.5em minus 0.4em\relax ISCA, Jun.
  2018, pp. 147--154. [Online]. Available:
  \url{http://www.isca-speech.org/archive/Odyssey_2018/abstracts/63.html}
\BIBentrySTDinterwordspacing

\bibitem{cieri2004fisher}
C.~Cieri, D.~Miller, and K.~Walker, ``The {F}isher {C}orpus: a resource for the
  next generations of speech-to-text.''\hskip 1em plus 0.5em minus 0.4em\relax
  Philadelphia: Linguistic Data Consortium, 2004.

\bibitem{Bredin2020}
H.~{Bredin}, R.~{Yin}, J.~M. {Coria}, G.~{Gelly}, P.~{Korshunov},
  M.~{Lavechin}, D.~{Fustes}, H.~{Titeux}, W.~{Bouaziz}, and M.-P. {Gill},
  ``{pyannote.audio: neural building blocks for speaker diarization},'' in
  \emph{ICASSP}, 2020.

\bibitem{huijbregts_speaker_2012}
\BIBentryALTinterwordspacing
M.~Huijbregts, D.~A. van Leeuwen, and C.~Wooters,
  ``\BIBforeignlanguage{en}{Speaker {Diarization} {Error} {Analysis} {Using}
  {Oracle} {Components}},'' \emph{\BIBforeignlanguage{en}{IEEE Transactions on
  Audio, Speech, and Language Processing}}, vol.~20, no.~2, pp. 393--403, Feb.
  2012. [Online]. Available: \url{http://ieeexplore.ieee.org/document/5955080/}
\BIBentrySTDinterwordspacing

\bibitem{maynard_conversation_1990}
S.~K. Maynard, ``\BIBforeignlanguage{en}{Conversation management in contrast:
  {Listener} response in {Japanese} and {American} {English}},''
  \emph{\BIBforeignlanguage{en}{Journal of Pragmatics}}, vol.~14, no.~3, pp.
  397--412, Jun. 1990.

\bibitem{Stivers10587}
\BIBentryALTinterwordspacing
T.~Stivers, N.~J. Enfield, P.~Brown, C.~Englert, M.~Hayashi, T.~Heinemann,
  G.~Hoymann, F.~Rossano, J.~P. de~Ruiter, K.-E. Yoon, and S.~C. Levinson,
  ``Universals and cultural variation in turn-taking in conversation,''
  \emph{Proceedings of the National Academy of Sciences}, vol. 106, no.~26, pp.
  10\,587--10\,592, 2009. [Online]. Available:
  \url{https://www.pnas.org/content/106/26/10587}
\BIBentrySTDinterwordspacing

\bibitem{liang_lu_effect_2009}
\BIBentryALTinterwordspacing
{Liang Lu}, {Yuan Dong}, {Xianyu Zhao}, {Jiqing Liu}, and {Haila Wang},
  ``\BIBforeignlanguage{en}{The effect of language factors for robust speaker
  recognition},'' in \emph{\BIBforeignlanguage{en}{ICASSP}}, 2009. [Online].
  Available: \url{http://ieeexplore.ieee.org/document/4960559/}
\BIBentrySTDinterwordspacing

\bibitem{xia_cross-lingual_2019}
\BIBentryALTinterwordspacing
W.~Xia, J.~Huang, and J.~H. Hansen, ``\BIBforeignlanguage{en}{Cross-lingual
  {Text}-independent {Speaker} {Verification} {Using} {Unsupervised}
  {Adversarial} {Discriminative} {Domain} {Adaptation}},'' in
  \emph{\BIBforeignlanguage{en}{{ICASSP}}}, 2019. [Online]. Available:
  \url{https://ieeexplore.ieee.org/document/8682259/}
\BIBentrySTDinterwordspacing

\bibitem{snyder2018xvector}
\BIBentryALTinterwordspacing
D.~Snyder, D.~Garcia-Romero, G.~Sell, D.~, and S.~Khudanpur, ``X-vectors:
  Robust {DNN} embeddings for speaker recognition,'' in \emph{ICASSP}, 2018.
  [Online]. Available:
  \url{http://www.danielpovey.com/files/2018_icassp_xvectors.pdf}
\BIBentrySTDinterwordspacing

\bibitem{burkner2018}
P.-C. Bürkner, ``Advanced {Bayesian} multilevel modeling with the {R} package
  {brms},'' \emph{The R Journal}, vol.~10, no.~1, pp. 395--411, 2018.

\bibitem{JSSv076i01}
\BIBentryALTinterwordspacing
B.~Carpenter, A.~Gelman, M.~Hoffman, D.~Lee, B.~Goodrich, M.~Betancourt,
  M.~Brubaker, J.~Guo, P.~Li, and A.~Riddell, ``Stan: A probabilistic
  programming language,'' \emph{Journal of Statistical Software, Articles},
  vol.~76, no.~1, pp. 1--32, 2017. [Online]. Available:
  \url{https://www.jstatsoft.org/v076/i01}
\BIBentrySTDinterwordspacing

\bibitem{lenth2020}
\BIBentryALTinterwordspacing
R.~Lenth, \emph{emmeans: Estimated Marginal Means, aka Least-Squares Means},
  2020, r package version 1.4.7. [Online]. Available:
  \url{https://CRAN.R-project.org/package=emmeans}
\BIBentrySTDinterwordspacing

\bibitem{zhang_fully_2019}
\BIBentryALTinterwordspacing
A.~Zhang, Q.~Wang, Z.~Zhu, J.~Paisley, and C.~Wang,
  ``\BIBforeignlanguage{en}{Fully {Supervised} {Speaker} {Diarization}},''
  \emph{\BIBforeignlanguage{en}{arXiv:1810.04719 [cs, eess, stat]}}, Feb. 2019,
  arXiv: 1810.04719. [Online]. Available: \url{http://arxiv.org/abs/1810.04719}
\BIBentrySTDinterwordspacing

\bibitem{fujita_end--end_2019}
\BIBentryALTinterwordspacing
Y.~Fujita, N.~Kanda, S.~Horiguchi, K.~Nagamatsu, and S.~Watanabe,
  ``\BIBforeignlanguage{en}{End-to-{End} {Neural} {Speaker} {Diarization} with
  {Permutation}-{Free} {Objectives}},''
  \emph{\BIBforeignlanguage{en}{Interspeech 2019}}, pp. 4300--4304, Sep. 2019,
  arXiv: 1909.05952. [Online]. Available: \url{http://arxiv.org/abs/1909.05952}
\BIBentrySTDinterwordspacing

\bibitem{fujita_end--end_20192}
\BIBentryALTinterwordspacing
Y.~Fujita, N.~Kanda, S.~Horiguchi, Y.~Xue, K.~Nagamatsu, and S.~Watanabe,
  ``\BIBforeignlanguage{en}{End-to-{End} {Neural} {Speaker} {Diarization} with
  {Self}-attention},'' \emph{\BIBforeignlanguage{en}{arXiv:1909.06247 [cs,
  eess]}}, Sep. 2019, arXiv: 1909.06247. [Online]. Available:
  \url{http://arxiv.org/abs/1909.06247}
\BIBentrySTDinterwordspacing

\bibitem{huijbregts_blame_2007}
M.~Huijbregts and C.~Wooters, ``\BIBforeignlanguage{en}{The {Blame} {Game}:
  {Performance} {Analysis} of {Speaker} {Diarization} {System} {Components}},''
  \emph{\BIBforeignlanguage{en}{Interspeech}}, 2007.

\end{thebibliography}

\end{document}